# Epitaxial growth and magnetic properties of kagome metal FeSn/elemental ferromagnet heterostructures

*Prajwal M. Laxmeesha[1], Tessa D. Tucker[1], Rajeev Kumar Rai[2], Shuchen Li[3], Myoung-Woo Yoo[3], Eric A. Stach[2], Axel Hoffmann[3,4], Steven J. May[1,*]*

[1] Department of Materials Science and Engineering, Drexel University, Philadelphia, USA

[2] Department of Materials Science and Engineering, University of Pennsylvania, Philadelphia, USA

[3] Department of Materials Science and Engineering, University of Illinois Urbana-Champaign, Urbana, USA

[4] Materials Research Laboratory, University of Illinois Urbana-Champaign, Urbana, USA

* smay@drexel.edu

**Abstract**

Binary kagome compounds $T_mX_n$ ($T$ = Mn, Fe, Co; $X$ = Sn, Ge; $m$:$n$ = 3:1, 3:2, 1:1) have garnered recent interest owing to the presence of both topological band crossings and flat bands arising from the geometry of the metal-site kagome lattice. To exploit these electronic features for potential applications in spintronics, the growth of high quality heterostructures is required. Here we report the synthesis of Fe/FeSn and Co/FeSn bilayers on $Al_2O_3$ substrates using molecular beam epitaxy to realize heterointerfaces between elemental ferromagnetic metals and antiferromagnetic kagome metals. Structural characterization using high-resolution X-ray diffraction, reflection high-energy electron diffraction, and electron microscopy reveals the FeSn films are flat and epitaxial. Rutherford backscattering spectroscopy was used to confirm the stoichiometric window where the FeSn phase is stabilized, while transport and magnetometry measurements were conducted to verify metallicity and magnetic ordering in the films. Exchange bias was observed, confirming the presence of antiferromagnetic order in the FeSn layers, paving the way for future studies of magnetism in kagome heterostructures and potential integration of these materials into devices.

**INTRODUCTION**

Topological semimetals have generated significant interest over the last decade. An emerging family of topological materials are binary kagome metals $T_mX_n$ ($T$ = Fe, Co, Mn; $X$ = Sn, Ge; $m$:$n$ = 1:1, 3:2, 3:1), which host both topological band features and flat bands near the Fermi level, giving rise to strong anisotropic anomalous and spin Hall effects [1-10]. Of these, FeSn is considered a prototypical kagome metal, which crystallizes in the hexagonal P6/*mmm* structure with lattice parameters $a$ = 5.297 Å and $c$ = 4.448 Å [11,12]. The structure follows an ABAB stacking sequence where the basal A plane is comprised of a network of Fe atoms that make up a kagome pattern with a Sn atom occupying the center of the resulting hexagon, yielding a planar composition of $Fe_3Sn$ as shown in Fig. 1(a). Within each kagome layer, the Fe spins couple ferromagnetically. The B plane is made up of monolayer stanene, $Sn_2$, resulting in an overall composition of FeSn. The neighboring $Fe_3Sn$ planes are coupled antiferromagnetically, giving rise to A-type antiferromagnetism with a bulk Néel temperature of 365 K [13,14]. Angle-resolved photoemission spectroscopy studies on bulk FeSn crystals have revealed the presence of Dirac fermions in surface and bulk states as well as the existence of flat bands below the Fermi level [15,16]. Recently, flat bands were also discovered in FeSn thin films originating from its surface kagome layers [17]. These authors predicted that the surface flat band in FeSn may generate novel spin-orbit-torques (SOT) when coupled with ferromagnets (FM) due to a strong contribution from the Berry curvature arising from the surface state.

To draw on these exotic SOTs and to use them effectively in devices, it is essential to synthesize high-quality crystalline thin films and interface them with other magnetic materials. Molecular beam epitaxy (MBE) has been employed previously to grow single-phase FeSn films directly on insulating perovskite oxide substrates, although realizing continuous films is a challenge. For instance, FeSn films grown on (111)-oriented perovskite $LaAlO_3$ and $SrTiO_3$ substrates typically crystallize as discreet 3D islands with a lateral feature size of ~20-200 nm [18,19], while continuous films grown on $SrTiO_3$ (111) have also been reported [20]. Epitaxial growth of FeSn has also been achieved with a combination of Pt and Ru as nonmagnetic metal buffer layers using magnetron sputtering [21]. While these buffer layers provide an effective template for epitaxy, Pt is known to exhibit strong spin-orbit coupling which may complicate efforts to isolate and quantify SOT signatures generated by FeSn and similar topological materials [22]. In this work, we show that continuous and epitaxial FeSn films can be grown on Co and Fe underlayers using MBE. A cartoon schematic of the stack is shown in Fig. 1(b). The ferromagnetic (FM) metals act as a buffer layer to facilitate uniform FeSn films on $Al_2O_3$ substrates, as revealed by electron microscopy, reflection high energy electron diffraction, and x-ray scattering. Exchange bias is observed in the Fe/FeSn and Co/FeSn bilayer junctions confirming the presence of exchange interactions between the elemental ferromagnets and the antiferromagnetic FeSn films.

**EXPERIMENTAL METHODS**

Heterostructures of Fe/FeSn and Co/FeSn were grown *via* MBE (Omicron modified LAB-10 system, base pressure ~5 × $10^{-10}$ Torr) on (0001)-oriented $Al_2O_3$ (MTI Corp.) through sublimation of Fe (99.95%, slug, Alfa Aesar) and Co (99.95%, shots, Alfa Aesar) and evaporation of Sn (99.9999%, shots, Alfa Aesar) from Knudsen cells. The corresponding cells were heated to ~ 1175 °C and ~1400 °C for Fe and Co buffer layer deposition, respectively. During FeSn deposition, the Fe and Sn cells were maintained at ~1125 °C and ~1040 °C, respectively, with co-deposition utilized to obtain FeSn. Film compositions were calibrated using a quartz crystal microbalance and Rutherford backscattering spectrometry (RBS). Prior to deposition, the substrates were sonicated in an acetone bath for 15 minutes, loaded into the chamber and heated to a temperature of ~ 425°C pre-deposition. The FM layer was deposited on $Al_2O_3$ (0001) substrates at ~400°C for Fe and ~ 450°C for Co and then the sample was cooled immediately to room temperature. Next, the Fe and Sn cell temperatures were adjusted to obtain ~1:1 stoichiometry and FeSn layer deposition was carried out at ~450°C. The sample was cooled following deposition without any post-growth annealing. The film quality was monitored *in situ* through reflection high-energy electron diffraction (RHEED) with an operating voltage of 14.5 kV. X-ray diffraction (XRD) and reflectivity (XRR) were measured using a Rigaku SmartLab diffractometer and analyzed using the software program GenX 3.6 to probe crystallinity, thickness, and interface/surface roughness. RBS was carried out at the Materials Research Laboratory, University of Illinois Urbana-Champaign and the Laboratory for Surface Modification at Rutgers University and analyzed using SIMNRA software package. Scanning electron microscopy (SEM) was performed using a Zeiss Supra 50VP field-emission SEM operated at 14 kV. High-angle annular dark-field (HAADF)-scanning transmission electron microscopy (STEM) with energy dispersive X-ray spectroscopy (EDS) was performed using an aberration-corrected JEOL NEOARM operating at 200 kV. The images and spectra were recorded using a 2-cm camera length and 27-mrad convergence angle. Transport measurements and magnetometry were carried out using a Physical Properties Measurement System (Quantum Design) with a vibrating sample magnetometry attachment. The diamagnetic signal from the substrate was not subtracted as the magnetization data was dominated by the Fe/Co ferromagnetic signal at the measurement fields used.

**RESULTS AND DISCUSSION**

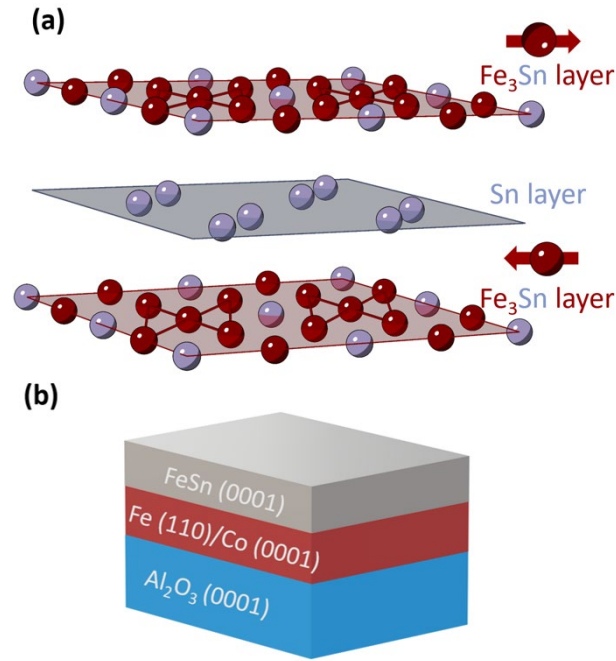

**FIG. 1.** (a) Layered structure of FeSn consisting of ferromagnetically ordered $Fe_3Sn$ layers that are antiferromagnetically coupling across monolayer stanene. (b) Schematic of bilayer FM/FeSn heterostructures grown on $Al_2O_3$ (0001) substrates.

Previous attempts to grow kagome metals directly on oxide substrates have often resulted in island growth and non-continuous films [18,19,23]. While Pt and Ru buffer layers have been employed to mitigate the propensity for island growth [21], we use Fe and Co as buffer layers, as both materials are also ferromagnetic. These $3d$ transition metals have been grown epitaxially on oxide substrates like MgO and $Al_2O_3$ [24-28]. For instance, Shiratsuchi and coauthors showed that under certain conditions epitaxial Fe (110) can grow as quasi-2D continuous films on $Al_2O_3$ (0001) substrates [25]. Co (0001) films on $Al_2O_3$ (0001) have also been grown epitaxially, despite a large mismatch of ~9% between them [27]. Prior to the deposition of our bilayer films, the FeSn stoichiometry was calibrated by altering the MBE fluxes and measuring the film composition of monolithic FeSn films grown directly on MgO (111) substrates using RBS, yielding general compositions of $Fe_{1+x}Sn_{1-x}$ ($x = 0 – 0.07$). The calibration films became more iron-rich in a linear fashion with increasing Fe flux as expected. The flux rates from these calibration growths were then used in the synthesis of the bilayer FM/FeSn structures. Interestingly, the bilayer films with different stoichiometries had noticeably different surface morphology. Supplementary Figure S1 shows electron micrographs from Fe/FeSn films with nominal compositions of $Fe_{0.99}Sn_{1.01}$, $Fe_{1.03}Sn_{0.97}$ and $Fe_{1.07}Sn_{0.93}$ respectively, based on the calibration data obtained from the monolithic FeSn films on MgO. $Fe_{0.99}Sn_{1.01}$ films showed a large volume of granular particles that were 200 – 500 nm in size on the surface. Upon increasing the Fe content in the film, these particles decrease in quantity and at a composition of $Fe_{1.07}Sn_{0.93}$, the particles are largely absent from the surface. Based on these observations, and previous reports of Sn dewetting from films [29, 30], we hypothesize that these particles are comprised of Sn (or a phase that is predominately Sn) that is not incorporated into the films. Informed by these calibration depositions, all the following measurements were carried out on FeSn samples with an estimated composition of $Fe_{1.05\pm0.02}Sn_{0.95\pm0.02}$.

Figure 2 shows RHEED images obtained after the deposition of each layer juxtaposed with a schematic of their hypothesized surface atomic configuration to show the epitaxial structure of the bilayer films. Oxygen atoms on an O-terminated $Al_2O_3$ (0001) surface form an irregular hexagonal pattern where four sides have length $s_1 = 2.87$ Å and two sides have length $s_2 = 2.53$ Å. Fe atoms in the (110) plane also form an irregular hexagonal pattern with four sides of length $s_1 = 2.48$ Å and two sides of length $s_2 = 2.87$ Å, and in the Co (0001) plane, the atoms form a regular hexagon with sides of length $s = 2.51$ Å. Similarly, the atoms in FeSn (0001) plane form a regular hexagon with sides of length $s = 2.65$ Å. These repeating hexagons stack on top of each other leading to epitaxial growth of Fe/Co and FeSn on $Al_2O_3$. Images of RHEED patterns obtained after growth of each layer show clear streaks indicating that the surfaces are smooth. Additionally, the same patterns from the FeSn layer were observed with every 60° in-plane rotation suggesting a sixfold rotational symmetry as expected from hexagonal lattice.

The lattice parameters and epitaxial orientation of the films were obtained from x-ray diffraction (XRD). Figure 3(a,b) shows $2\theta$-$\omega$ XRD scans from three Fe/FeSn (F1, F2 and F3) and Co/FeSn (C1, C2 and C3) bilayers of varying FeSn thicknesses. The scattering vector was fixed perpendicular to the film surface, and excluding the substrate peaks, only 000$l$, $hh$0 and 000$l$ reflections were observed from FeSn, Fe and Co, respectively. The presence of significant Laue oscillations about the Fe, Co and FeSn Bragg peaks points to sharp interfaces between each layer. The thickness fringes arising from the Co and Fe layers are convoluted with the FeSn 0002 Bragg peak positions, especially in films with thinner FeSn layers. To extract quantitative information from the diffraction data, we simulated the *2θ-ω* patterns from these heterostructures using GenX [31]. Through comparisons between the measured and simulated data, the thickness and *c*-axis parameters of each layer are determined. The parameters obtained for Fe ($d_{110}$ = 2.03 Å, corresponding to $a$ = 2.87 Å) and Co ($c$ = 4.07 Å) were within 0.1% and 0.25% of bulk values [32]. The average *c*-axis parameter for the FeSn layer in Fe/FeSn films is 4.44 ± 0.01 Å and in Co/FeSn films is 4.40 ± 0.03 Å. For comparison, the *c*-axis length for bulk FeSn is 4.448 Å [11]. The origin of these deviations from bulk *c*-axis values is likely due to the small degree of off-stoichiometry within the films. Noteworthily, in Co/FeSn films, the obtained lattice parameters $a_{FeSn}$ and $c_{FeSn}$ of sub-5 nm films were different than of 40 nm films: $a_{4nm}$ = 5.402 Å and $a_{40nm}$ = 5.429 Å, whereas $c_{4nm}$ = 4.37 Å and $c_{40nm}$ = 4.42 Å. We attribute the decreased *a*-axis and *c*-axis parameter in ultrathin FeSn films on Co to chemical intermixing between FeSn and Co layers leading to Co incorporation in FeSn near the interface. As Meier and coauthors have shown [13], in $Fe_{1-x}Co_xSn$ both *a*-axis and *c*-axis parameters decrease with increasing Co concentration. In contrast, we observe minimal change in lattice parameters for 8 nm and 40 nm FeSn films on Fe.

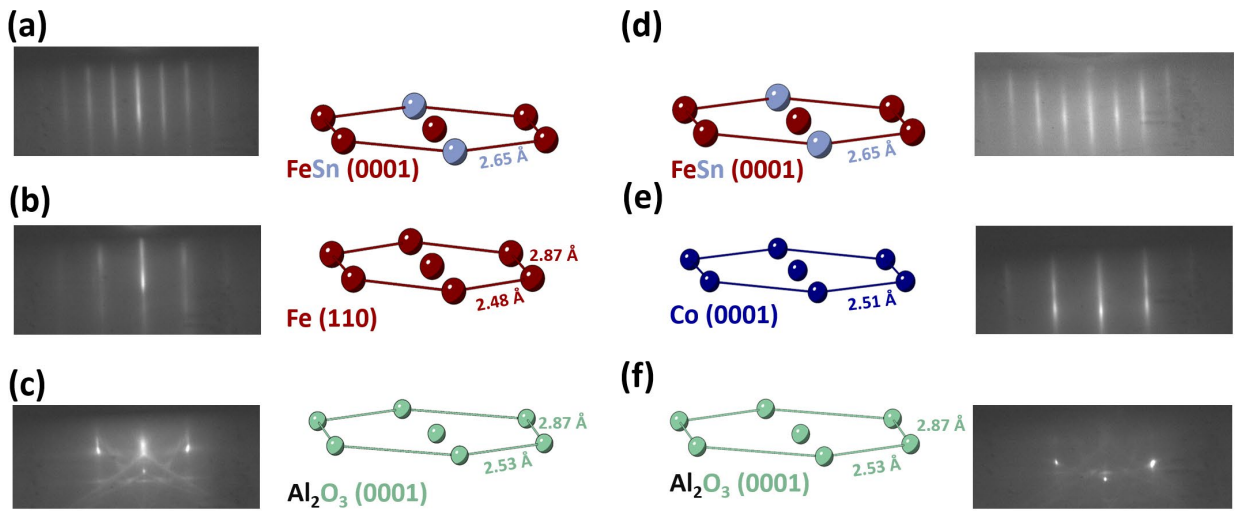

**FIG. 2**. RHEED images obtained during growth along with schematics of the hexagonal-like structural motifs from (a) FeSn layer on Fe, (b) Fe on $Al_2O_3$, (c) $Al_2O_3$ substrate, (d) FeSn layer on Co, (e) Co on $Al_2O_3$, and (f) $Al_2O_3$ substrate.

To determine if the films are epitaxial, in-plane diffraction measurements were performed. $\phi$ scans along a fixed $\chi$ angle, shown in Fig. 3(e, f), reveal six distinct in-plane $2\bar{1}\bar{1}0$ Bragg peaks from the FeSn layer in Fe/FeSn and Co/FeSn respectively. The peaks are evenly spaced 60° apart from each other and are spaced ~30° (±5°) apart from the six $2\bar{1}\bar{1}0$ Bragg peaks of $Al_2O_3$. As shown schematically in Supplementary Fig. S2, this 30° rotation provides better crystallographic alignment between $Al_2O_3$ (0001) and FeSn (0001). A small deviation, ~5°, is observed between the ideal 30° rotation, the origin of which is unknown but likely arises from the presence of the Fe or Co buffer layer. This confirms the epitaxial nature of the FeSn layers with respect to the substrate, and by correlation, likely the ferromagnetic layers Fe and Co as well. The epitaxial relationship is as follows: $Al_2O_3$ [0001] || FeSn [0001] and $Al_2O_3$ [$2\bar{1}\bar{1}0$] || FeSn [$10\bar{1}0$]. Due to the small thicknesses of the ferromagnetic layers, no in-plane Bragg peaks were observed from either Co or Fe. The full-width-at-half-maximum (FWHM) of the FeSn $2\bar{1}\bar{1}0$ peaks from the film grown on Co is 1.9°, which is smaller than that of the film on Fe at 6.9°, indicating a better degree of in-plane crystallinity for the film grown on the Co buffer. This is likely due to the 5% mismatch between Co and FeSn hexagons compared to 7% between Fe and FeSn hexagons, as well as the better symmetry matching between FeSn and the Co (0001) plane compared to the Fe (110) plane. However, this is not reflected in the rocking curve measurements from the 0002 FeSn peaks as both Fe/FeSn and Co/FeSn films have comparable FWHM values of 0.09° and 0.07°, respectively, for FeSn films ~10 nm in thicknesses (Fig. S3). For comparison, the $Al_2O_3$ 0006 rocking curve had a FWHM of 0.03° under the same measurement conditions, which points to both films having some degree of mosaicity. As the FeSn film thickness is increased, the rocking curve FWHM values also increase with 40-nm thick films exhibiting FWHM on the order of 0.2°. Supplementary Figure S3 shows a compilation of the FWHM of films plotted against their thicknesses and a clear trend is observed, indicating that the film crystalline quality declines slightly at larger thicknesses. While the films are epitaxial, they do not appear to be coherently strained to the $Al_2O_3$ substrate as the RHEED patterns show clear differences in $k_x$ spacings between the specular and scattered streaks of each layer within the heterostructure (Fig. S4).

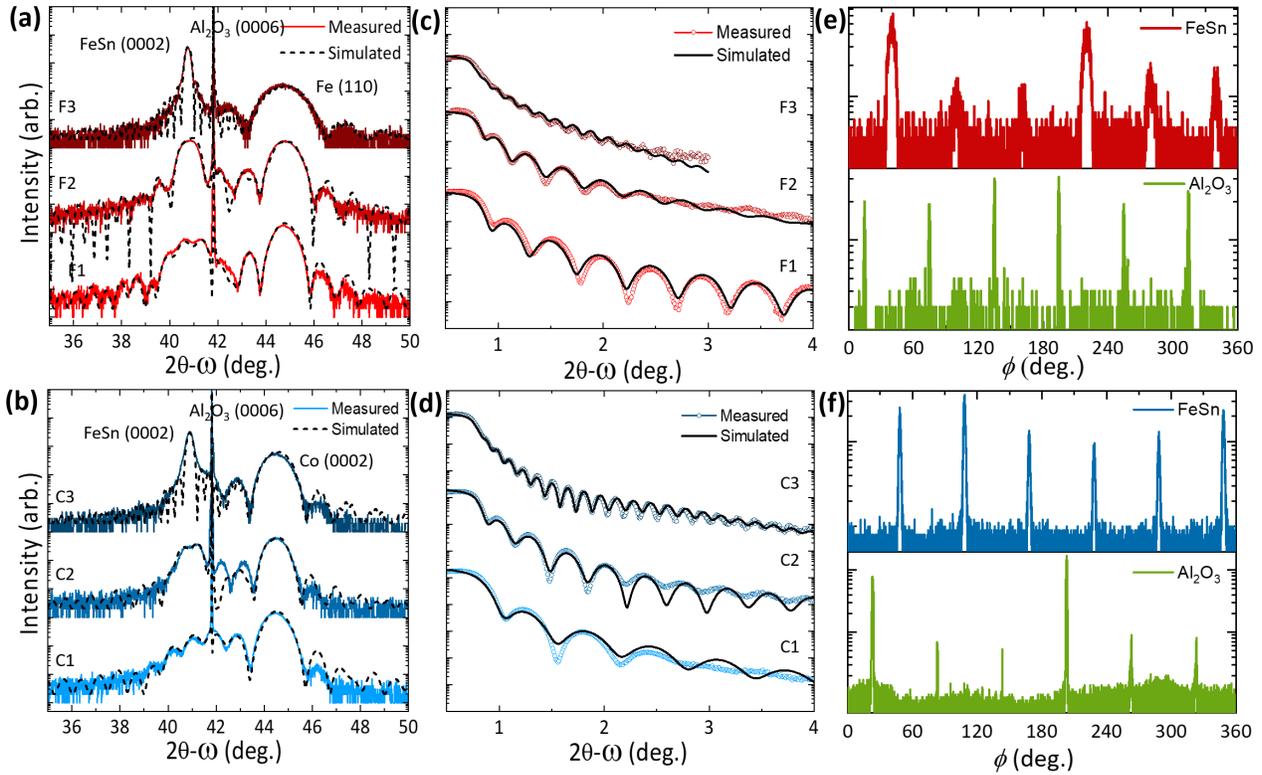

**FIG 3.** X-ray structural characterization of Fe/FeSn and Co/FeSn heterostructures. Fe-based bilayers are plotted in red and Co-based bilayers are plotted in blue. (a,b) $2\theta$-$\omega$ scans and simulations from heterostructures of varying FeSn thickness, (c,d) XRR scans and fits, (e,f) $\phi$ scans from bilayers with ~40 nm thick FeSn layers.

The bilayer thickness and roughness values were further examined using XRR on the same set of films. The roughness of each layer in all films was found to be less than 1 nm, while the obtained scattering length densities of all layers were within 5% of their theoretical values. There is good agreement between the thicknesses calculated from the XRD simulations and XRR fits (as shown in Supplementary Table S1) implying the films are crystalline throughout their entire thickness. It is worth noting that a ~1 nm thick $FeSnO_x$ passive oxide layer was used in the model to account for surface oxidation which led to better fits. Resistivity measurements from the bilayer samples exhibit metallic temperature dependence (as shown in Supplementary Fig. S5), a result that is consistent with laterally continuous films.

Cross-sectional STEM and EDS analyses were conducted to probe the interface in a representative bilayer Co/FeSn film. Figure 4 (a) shows a HAADF-STEM image along the $[10\bar{1}0]$ direction revealing two distinct layers atop the insulating $Al_2O_3$ substrate. In this sample, the Co and FeSn layers are ~7 nm and ~35 nm, respectively, featuring sharp interfaces between $Al_2O_3$-

Co and Co-FeSn, as illustrated in Supplementary Fig. S6. The high-resolution image of the FeSn layer shown in Fig. 4(b) confirms the expected ABAB stacking of the $Fe_3Sn$ and $Sn_2$ planes in FeSn. A schematic of the crystal structure along the $[10\bar{1}0]$ direction is shown in Fig. 4(c). To further understand the distribution of Al, O, Co, Fe and Sn within the heterostructure and to visualize the intermixing at the Co-FeSn interface, EDS spectra were acquired, and corresponding elemental maps are shown in Fig. 4(e). An EDS line profile along the green line drawn in Fig. 4(e) is shown in Fig. 4(d). The $Al_2O_3$-Co interface is very sharp with limited intermixing and minimal roughness. However, at the Co-FeSn interface, some intermixing of Co into FeSn is present in the first 4 nm of the FeSn layer. This observation falls in line with the *c*-axis shrinkage noticed from the XRD measurements of the ultrathin FeSn films on Co. Interestingly, an additional layer with enhanced contrast was observed in the high-resolution image of $Al_2O_3$-Co interface (Supplementary Fig. S6), which originates from Sn atoms, as confirmed by EDS. The Sn diffusion and accumulation at the interface doesn't occur by way of uniform diffusion of Sn through the Co layer but through pillar formation in the Co layer, as shown in the EDS map in Supplementary Fig. S6.

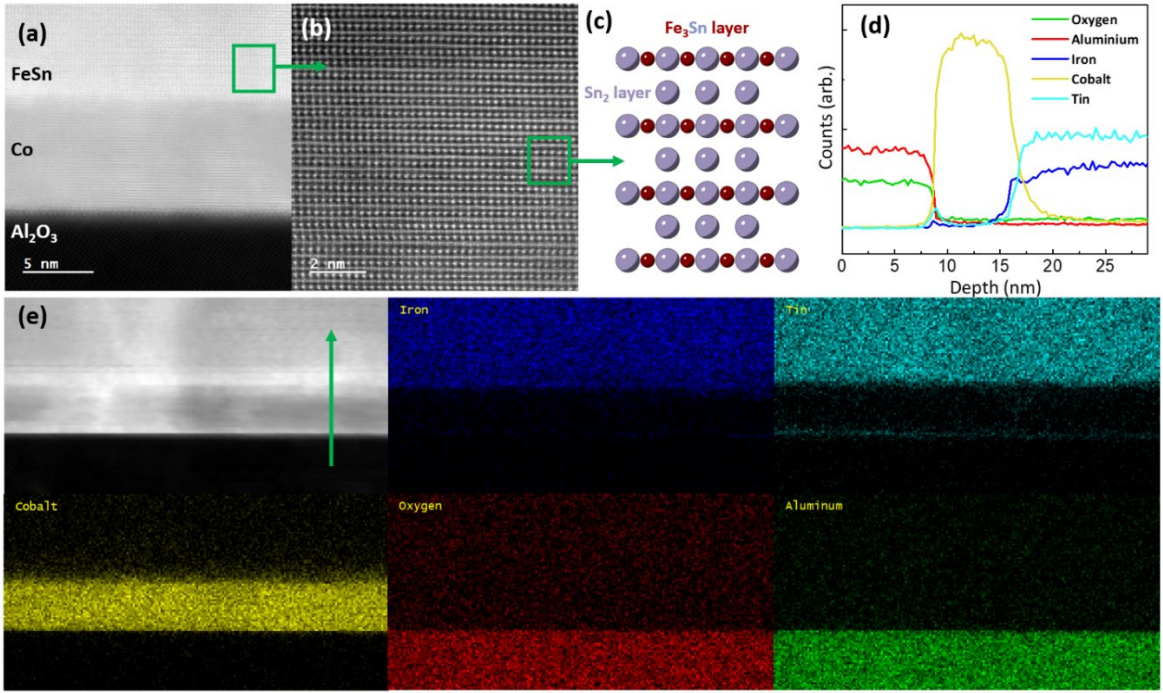

**FIG. 4.** Cross-sectional STEM-EDS of a Co/FeSn bilayer film. (a) HAADF-STEM image showing $Al_2O_3$-Co and Co-FeSn interfaces. (b) High-resolution magnified image of the FeSn layer along $[10\bar{1}0]$ highlighted in green. (c) Animated depiction of FeSn viewed along the $[10\bar{1}0]$ direction. (d) Line profile EDS map along the two interfaces. (e) EDS elemental distribution map of O, Al, Co, Sn and Fe in the heterostructure.

These narrow Sn pillars span the entire thickness of the Co layer and disperse at the interface with the $Al_2O_3$ substrate. Away from the pillars, the line profile and the EDS map both indicate that Sn is exclusively concentrated at the interface and is not dispersed within the Co layers. This, supported by the streaky nature of our RHEED patterns and significant Laue oscillations from the Bragg peaks, indicates that the films are smooth and continuous with well-defined interfaces.

We performed magnetometry on the Fe (7 nm)/ FeSn (38 nm) and Co (8 nm)/ FeSn (40 nm) samples, the results of which are displayed in Fig. 5. At 300 K, the saturation magnetization ($M_s$) for the two films were found to be ~1675 kA/m and ~1390 kA/m, and coercive field strength ($\mu_0 H_c$) to be 9.4 mT and 14.4 mT, respectively. These values are consistent with previous reports in literature for monolithic Fe and Co films [33, 34], indicating that the FeSn layers are not contributing a significant magnetization signal as expected for antiferromagnets. To elucidate the interaction between the ferromagnetic metals and antiferromagnetic FeSn layers, isothermal magnetization measurements as a function of applied field were conducted at various temperatures. The samples were field cooled from 400 K, which is well above $T_N$ for FeSn, under an in-plane applied field. Field scans were conducted twice at each temperature, once under positive field (2 T) cooling and again under negative field cooling (-2 T).

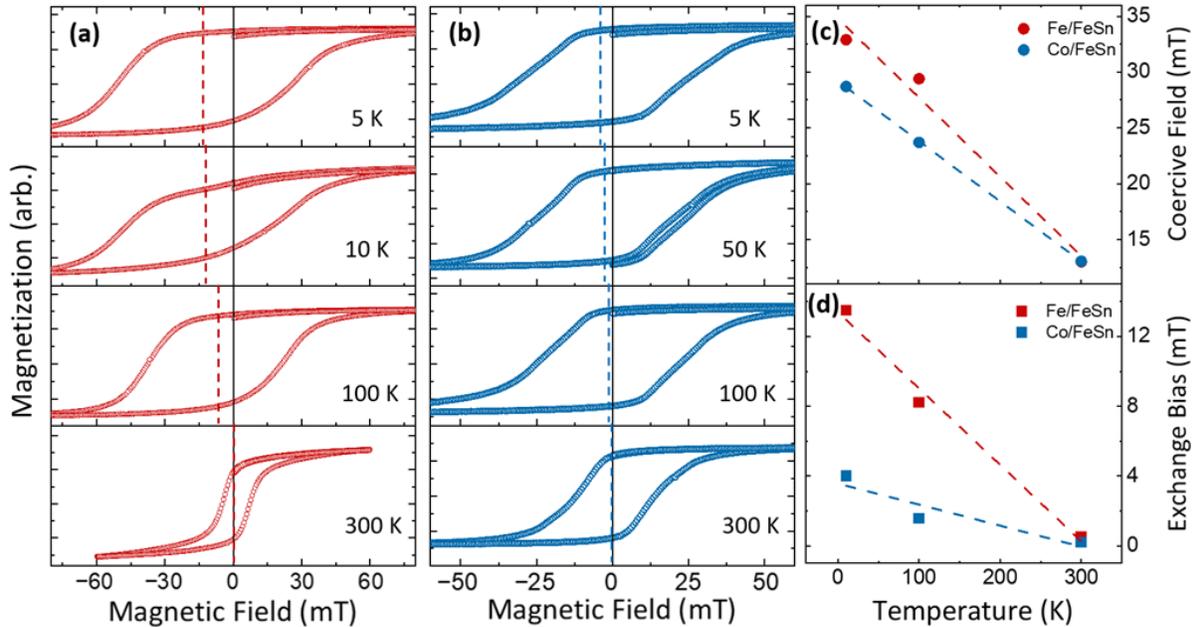

**FIG. 5.** Magnetization hysteresis loops measured at multiple temperatures from a) Fe/FeSn and b) Co/FeSn bilayers after field cooling under +2 T. Data was obtained after field cooling at 2 T from 400 K with the magnetic field applied in-plane.

Average exchange bias field and coercive field calculated after ±FC, plotted against temperature from c) Fe/FeSn and d) Co/FeSn. At 300 K, little-to-no loop shift was observed, however, at low temperatures, a negative exchange bias was observed when the samples were field-cooled under a positive field, as shown in Fig. 5(a) and 5(b) for both Fe/FeSn and Co/FeSn, respectively. A positive exchange bias was also observed when the samples were field cooled under a negative field as shown in Supplementary Fig. S7. The shift in hysteresis loops were accompanied by an increase in $H_C$ further pointing to an exchange interaction between the ferromagnetic Fe and Co layers and the antiferromagnetic FeSn layer [35]. Figure 5(c,d) shows the temperature dependence of exchange bias field ($H_{EB}$) and coercive fields ($H_C$) in both Fe and Co-based heterostructures. $H_{EB}$ was calculated by averaging the magnitude of the positive and negative exchange biases observed with ±2 T field cooling. In both films, $H_{EB}$ and $H_C$ increase with decreasing temperature. Both $H_C$ and $H_{EB}$ are larger in magnitude for Fe/FeSn than Co/FeSn. One possible explanation could be the intermixing at the Co-FeSn interface. In bulk $Fe_{1-x}Co_xSn$ random alloys, the ordered antiferromagnetic moments undergo a reorientation with substitution of Co in the lattice – from planar to tilted and then to axial antiferromagnetism. At $x = 1$ (CoSn) the crystal becomes paramagnetic [12]. Hence, it is possible the moments in the intermixed phase at the Co-FeSn interface are partially reoriented from planar to tilted/axial, leading to a decrease in exchange bias. The $H_C$ and $H_{EB}$ values in both Co/FeSn and Fe/FeSn are lower in comparison to the reported values for permalloy/FeSn [16].

**CONCLUSIONS**

We have demonstrated the growth of epitaxial and continuous thin films of FeSn by utilizing elemental ferromagnetic metal buffer layers, BCC Fe (110) and HCP Co (0001), on $Al_2O_3$ (0001) substrates. By growing films that are slightly Sn deficient, we show that heterostructures with smooth surfaces, well-defined interfaces, and a high degree of crystallinity can be obtained. The FeSn films are antiferromagnetic as confirmed by the presence of exchange bias in both Fe/FeSn and Co/FeSn bilayers. We predict that other typical ferromagnets with FCC and BCC structures like Ni and permalloy could also be used as a buffer layer to grow FeSn and other kagome crystals, as the (111) and (110) surfaces, respectively, offer the hexagonal imprint required for epitaxy of these materials. The ability to form epitaxial ferromagnetic heterostructures with topological kagome materials opens the door to studies of spin-orbit torques and spin transport in these systems.

**ACKNOWLEDGEMENTS**

The authors at Drexel and UIUC were supported by the National Science Foundation under Grant No. ECCS-2031870, except M.-W.Y., who was supported by the National Science Foundation through the University of Illinois Urbana-Champaign Materials Research Science and Engineering Center Grant No. DMR-1720633. R.K.R. and E.D.S. were supported by the National Science Foundation under Grant No. DMR-2122102. We would like to thank Hussein Hijazi at Rutgers University for performing timely RBS experiments and Dr. Craig Johnson of the Materials Characterization Core (MCC) at Drexel University for the assistance in preparing samples for and performing TEM. We also acknowledge the characterization facilities at MCC of the College of Engineering at Drexel University; the diffractometer was acquired with support from the National Science Foundation under Grant No. DMR-1040166. The acquisition of the PPMS used for transport and magnetometry measurements was supported by the U.S. Army Research Office under Grant No. W911NF-11-1-0283.

**SUPPLEMENTARY INFORMATION**

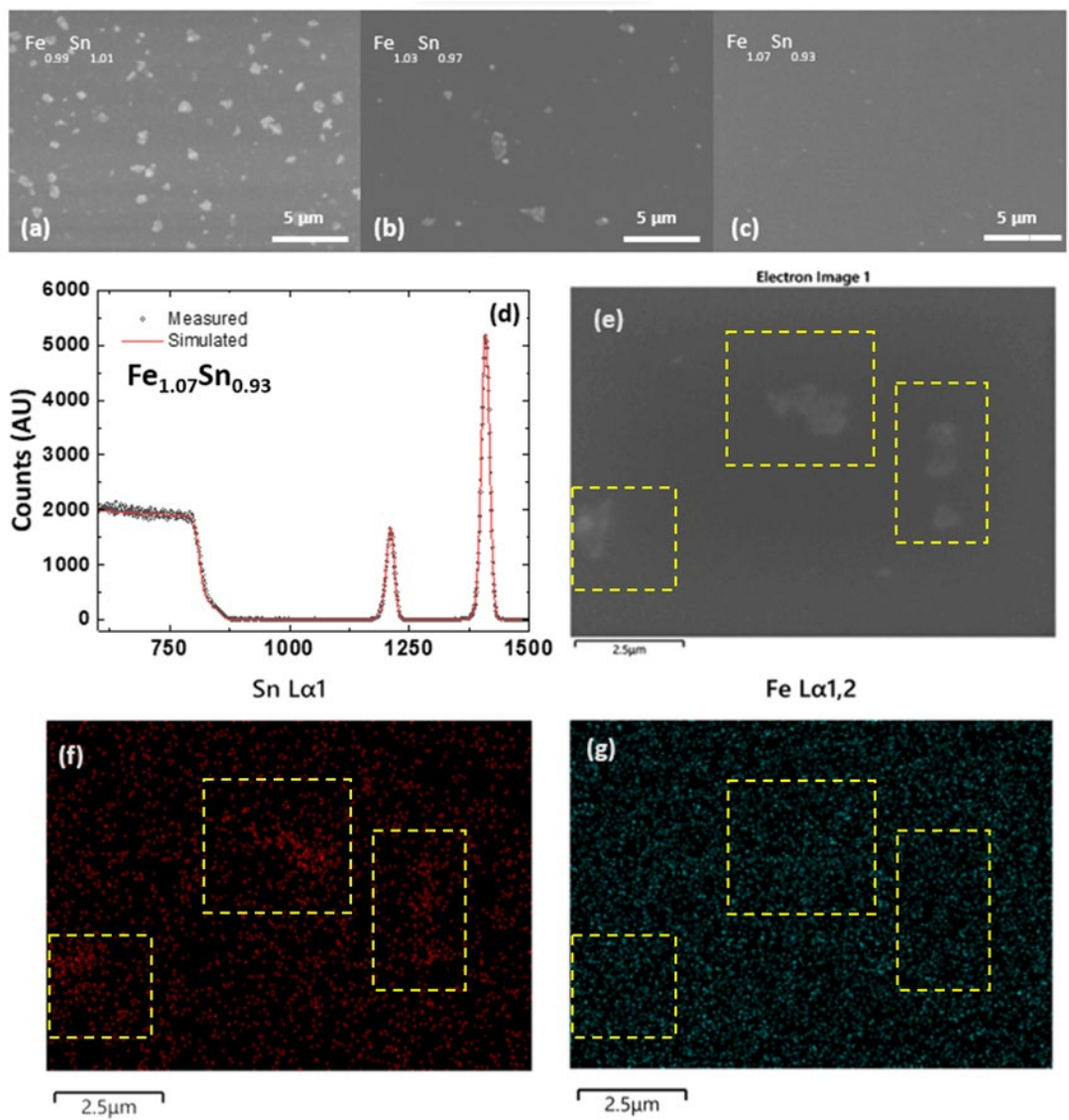

**FIG. S1.** Surface morphology of Fe/FeSn films measured using SEM. (a) $Fe_{0.99}Sn_{1.01}$ showing granular particles. (b) $Fe_{1.03}Sn_{0.97}$ showing a decrease in the number particles, and (c) $Fe_{1.07}Sn_{0.93}$ showing almost no particles on the surface. (d) Example of Rutherford backscattering spectrometry data and fit used to obtain stoichiometry from Fe-Sn films grown directly on MgO. (e) Electron beam image of the FeSn surface from a Fe/FeSn heterostructrue. Yellow boxes highlight surface particles. (f) Isolated Energy despersive spectroscopy (EDS) map of Sn showing higher Sn concentration corresponding to the surface particles observed in (e) within the boxes. (g) Uniform Fe distribution throughout the surface, and in the boxes indicating the surface particles are Sn rich.

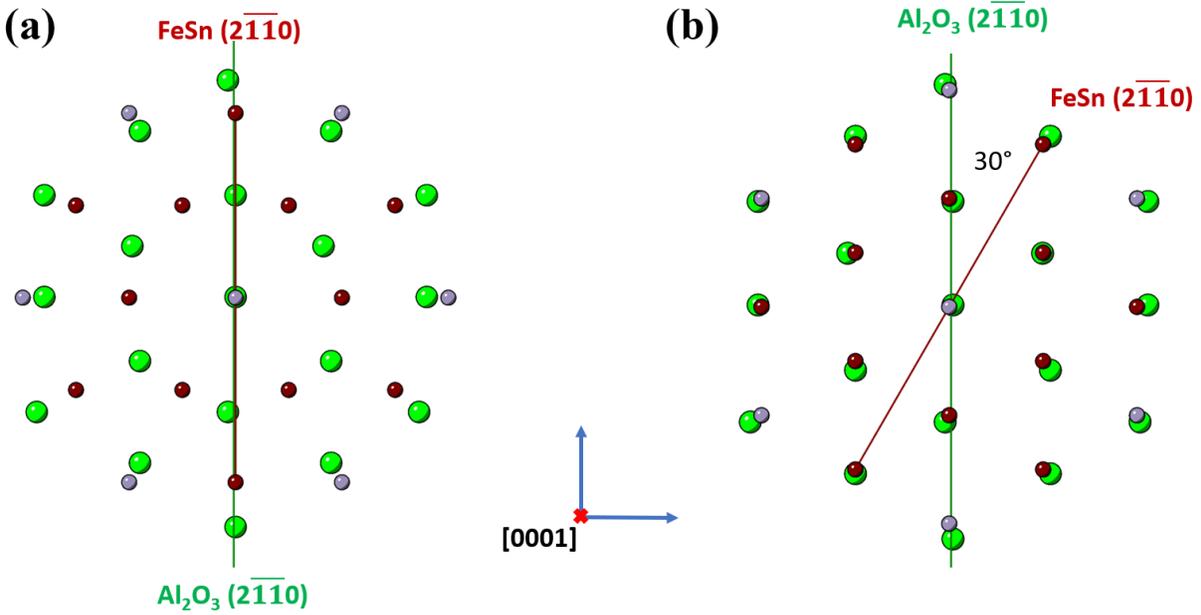

**FIG. S2.** Surface crystallographic configuration of FeSn and Al$_2$O$_3$ lattices viewed from the [0001] direction. (a) (2$\bar{1}\bar{1}$0) planes of both FeSn (red line) and Al$_2$O$_3$ (green line) are superimposed on one another. (b) By rotating the FeSn lattice by 30°, a more favorable alignment is achieved. This is also reflected in the $\phi$ scans as each of the six 2$\bar{1}\bar{1}$0 Bragg peaks of FeSn are separated by 30° from their Al$_2$O$_3$ counterparts.

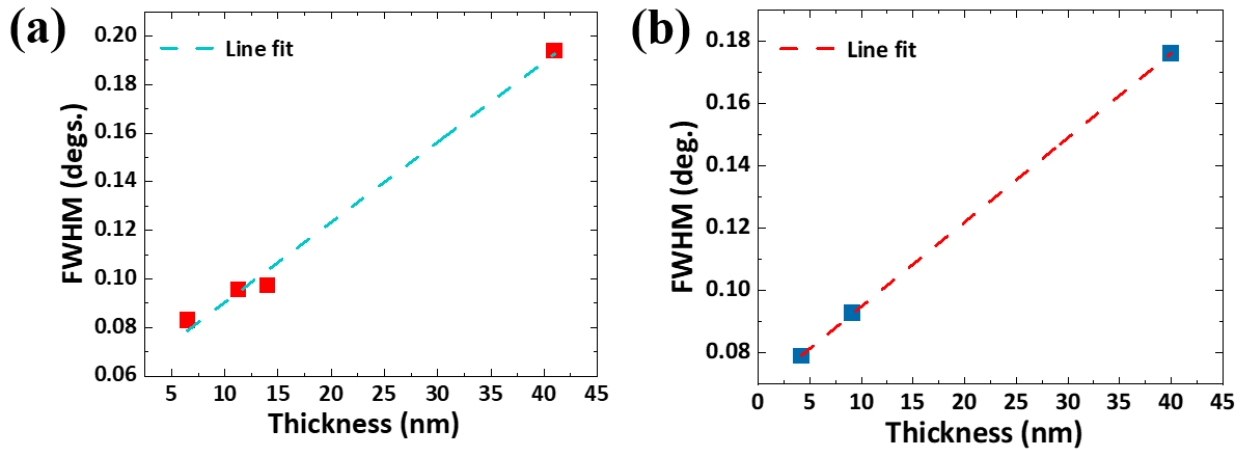

**FIG. S3.** Full-width-half-maximum of rocking curves measured at the FeSn 0002 Bragg peak plotted as a function of thickness for (a) Fe/FeSn films and (b) Co/FeSn films.

**Table S1.** Comparison of layer thicknesses obtained from x-ray diffraction (XRD) and x-ray reflectivity (XRR).

| Layer | $t_{XRD\ simulation}$ (nm) | | | $t_{XRR\ fit}$ (nm) | | |
|---|---|---|---|---|---|---|
| | F1 | F2 | F3 | F1 | F2 | F3 |
| FeSn | 7.2 | 11.5 | 42 | 7.5 | 12 | 41.2 |
| Fe | 9.3 | 8.5 | 5.9 | 9.5 | 9.5 | 8 |
| | C1 | C2 | C3 | C1 | C2 | C3 |
| FeSn | 8 | 4 | 42 | 9.1 | 4.3 | 40 |
| Co | 9.8 | 8.6 | 8.1 | 10.6 | 9 | 8.4 |

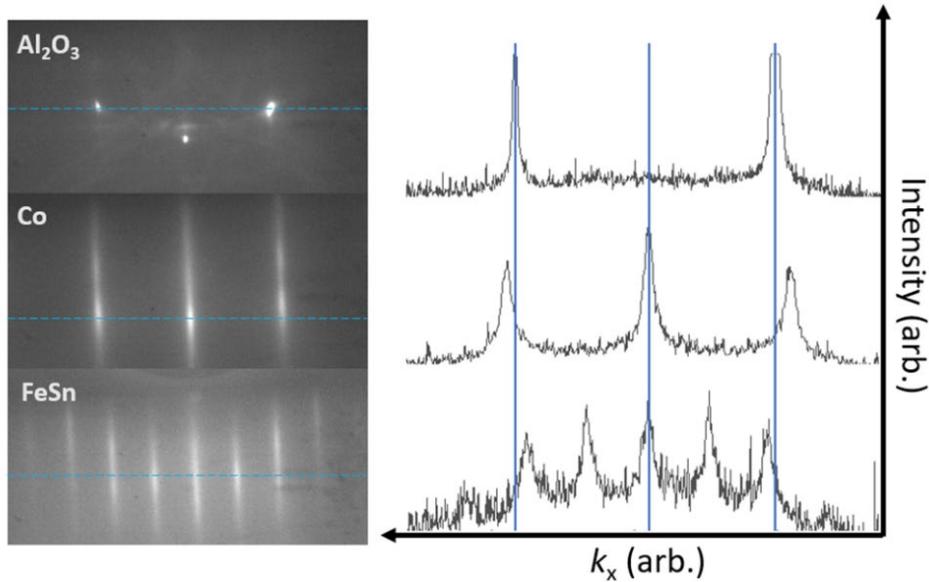

**FIG. S4**. Line scans through RHEED images obtained from a FeSn/Co/Al$_2$O$_3$ heterostructure. The difference between the streak spacing for each layer indicates a difference in in-plane lattice spacing, consistent with strain relaxation within each layer.

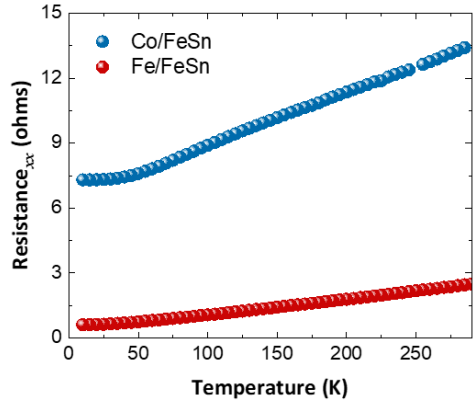

**FIG. S5.** Longitudinal electronic transport of Fe (7 nm)/ FeSn (38 nm) and Co (8 nm)/ FeSn (40 nm) bilayers showing metallic conductivity in both heterostructures.

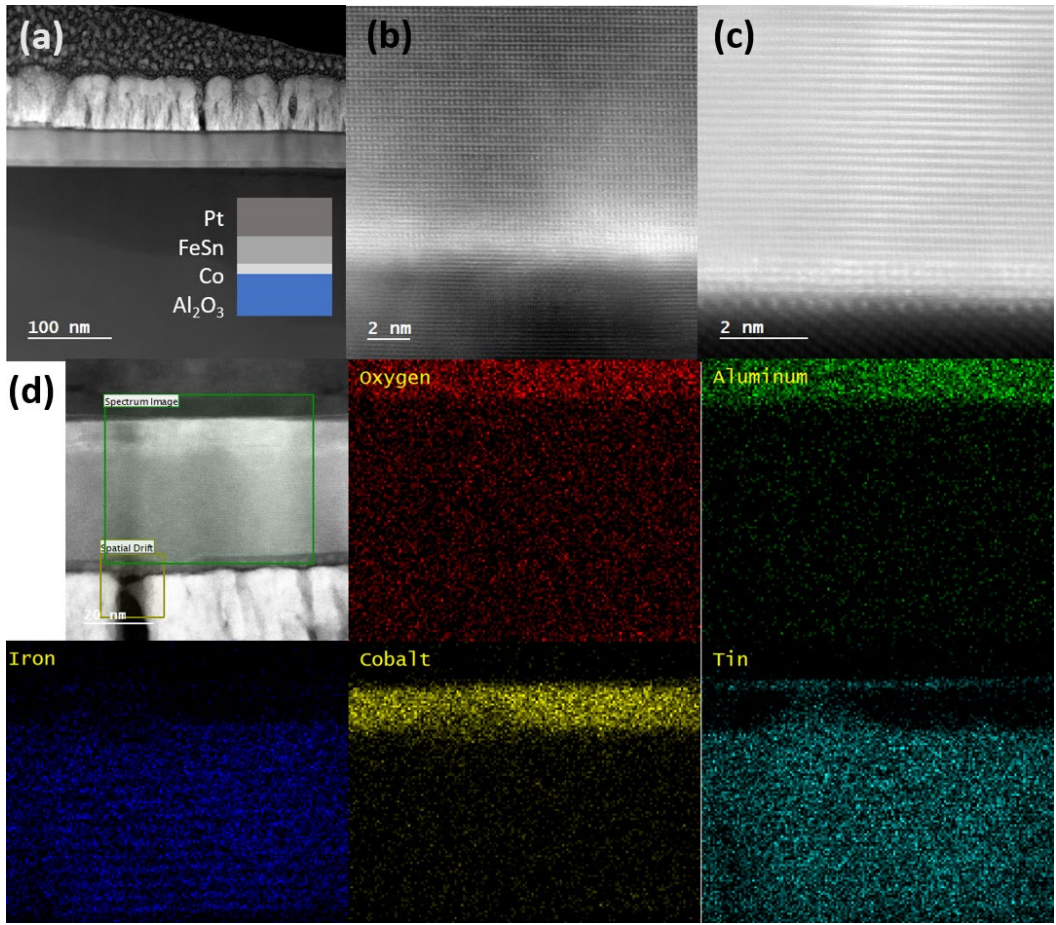

**FIG. S6.** (a) Low-magnification HAADF-STEM image of a Co/FeSn bilayer. Inset shows a schematic of the film stack with protective Pt coating. (b) HAADF-STEM image showing Co-FeSn interface, and (c) $Al_2O_3$-Co interface. (d) EDS map showing the formation of a Sn-rich channel in the Co layer.

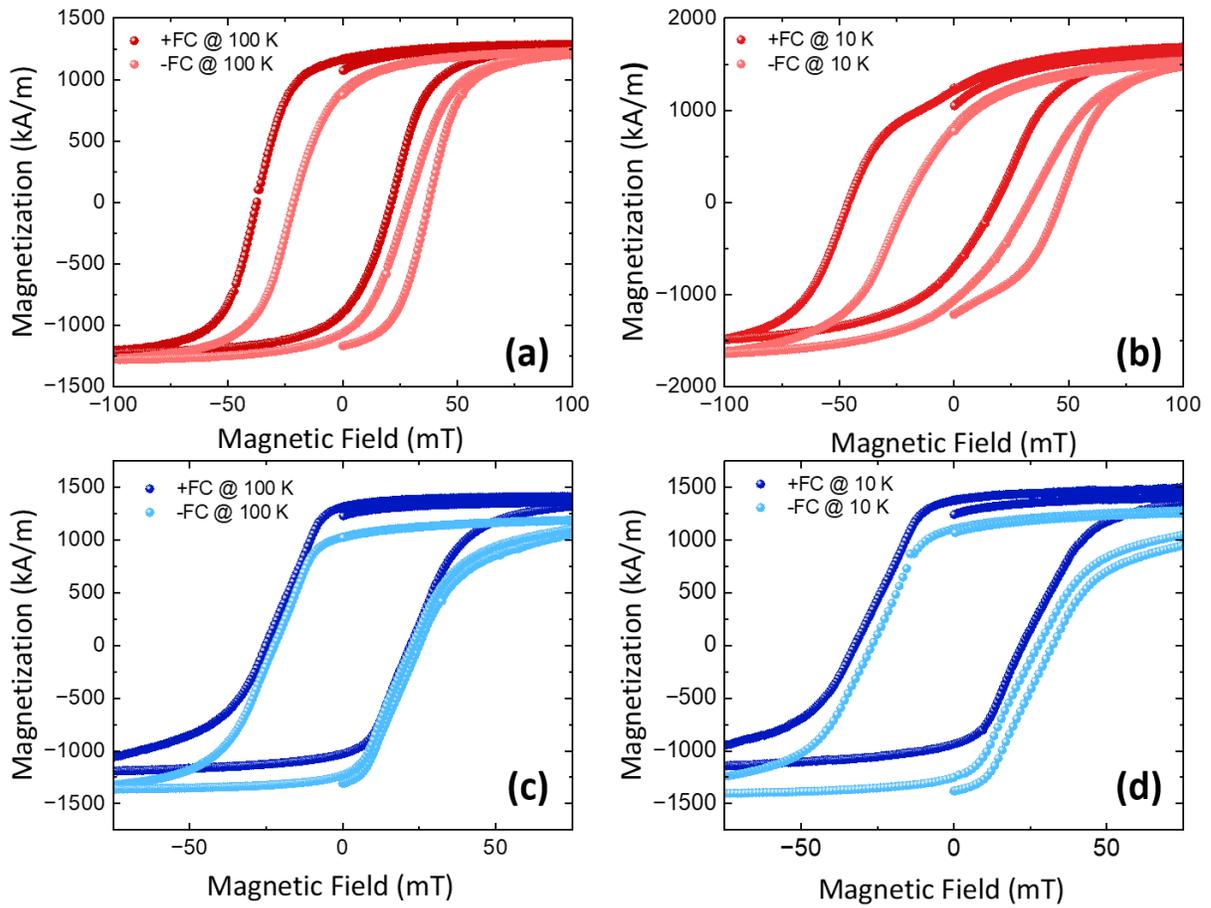

**FIG. S7.** Field scans after positive (dark shades) and negative (light shades) field cooling under ±2 T at multiple temperature. (a) Field scan from Fe/FeSn at 100 K. (b) Field scan from Fe/FeSn at 10 K. (c) Field scan from Co/FeSn at 100 K. (d) Field scan from Co/FeSn at 10 K.